\documentclass[journal=jpclcd,manuscript=letter]{achemso}

\usepackage{amsmath}
\usepackage{mathrsfs}
\usepackage{array}
\usepackage{graphicx}
\usepackage{amstext}
\usepackage{amsfonts}
\usepackage{bm}
\usepackage{amssymb}
\usepackage{diagbox}
\usepackage{natbib}
\usepackage[utf8x]{inputenc}

\usepackage{indentfirst}

\date{\today}

\author{Pei-Yun Yang}
\affiliation{Department of Chemistry, Massachusetts Institute of Technology, Massachusetts, 02139 USA }
\alsoaffiliation{Beijing Computational Science Research Center, Beijing 100193, Peoples Republic of China}
\author{Jianshu Cao}
\email{jianshu@mit.edu}
\affiliation{Department of Chemistry, Massachusetts Institute of Technology, Massachusetts, 02139 USA }

\title{Steady-State Analysis of Light-harvesting Energy Transfer Driven by Incoherent Light:  From Dimers to Networks}

\begin{document}

\begin{tocentry}
\centerline{\scalebox{0.35}{\includegraphics[trim=0 0 120 100,clip]{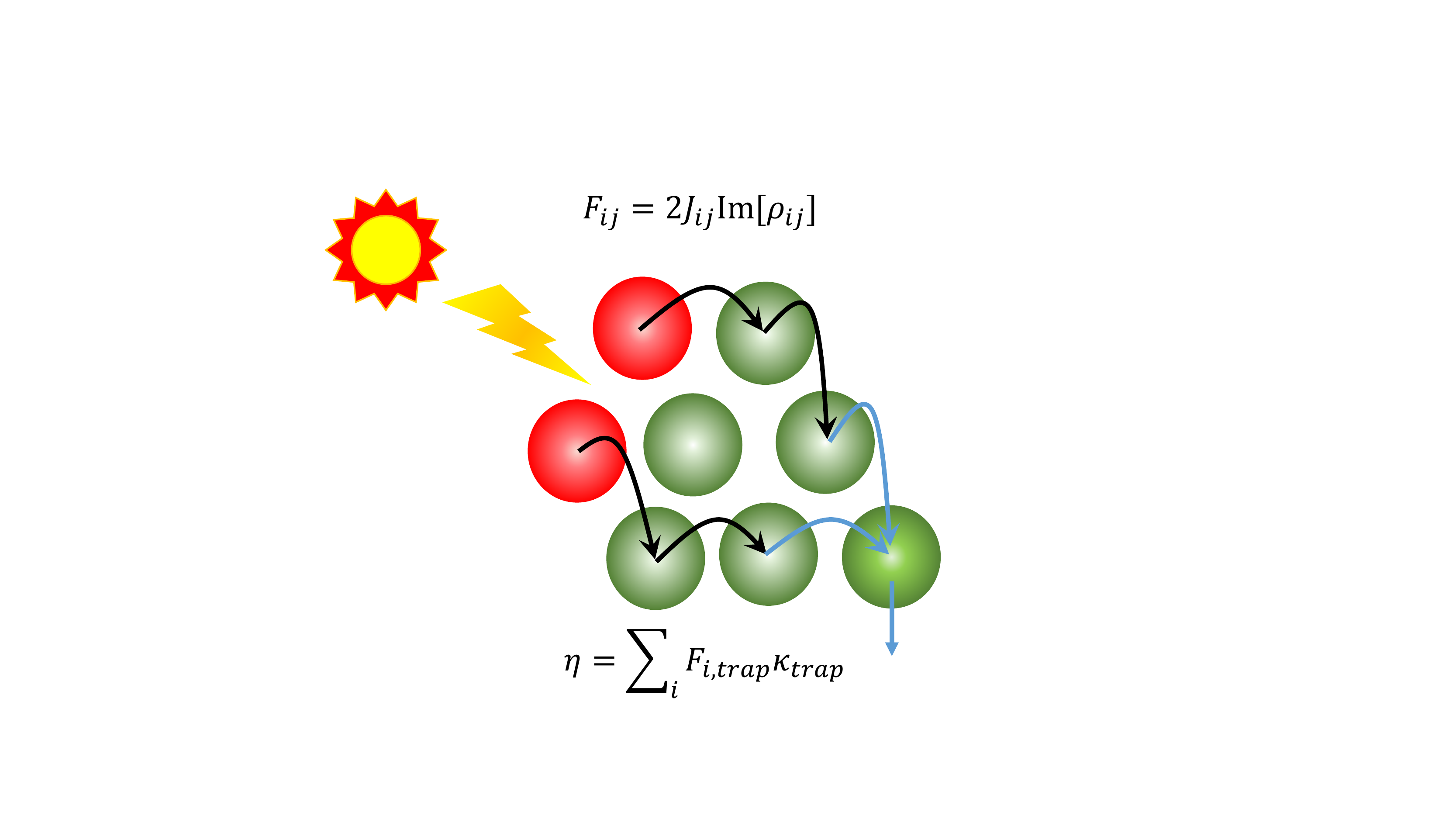}}}
\end{tocentry}

\begin{abstract}
The question of how quantum coherence facilitates energy transfer has been intensively debated in the scientific community.  Since natural and artificial light-harvesting units operate under the stationary condition, we address this question
via a non-equilibrium steady-state analysis of a molecular dimer irradiated by incoherent sunlight
and then generalize the key predictions to arbitrarily-complex exciton networks.
The central result of the steady-state analysis is the coherence-flux-efficiency relation:
$ \eta = c \sum_{i \neq j}F_{ij} \kappa_j = 2c \sum_{i \neq j} J_{ij} {\rm Im} [{\rho}_{ij}] \kappa_j $
with $ c $ the normalization constant.
In this relation, the first equality indicates that energy transfer efficiency $\eta$ is uniquely determined by the trapping flux, which is the product of
flux $F$ and branching ratio $\kappa$ for trapping at the reaction centers, and the second equality indicates that
the energy transfer flux $F$ is equivalent to quantum coherence
measured by the imaginary part of the off-diagonal density matrix, i.e., $ F_{ij}= 2J_{ij} {\rm Im} [{\rho}_{ij}] $.
Consequently, maximal steady-state coherence  gives rise to optimal efficiency.
The coherence-flux-efficiency relation holds rigorously and generally for any exciton networks of arbitrary connectivity
under the stationary condition and is not limited to incoherent radiation or incoherent pumping.
For light-harvesting systems under incoherent light,
non-equilibrium energy transfer flux (i.e. steady-state coherence) is
driven by the breakdown of detailed balance and by the quantum interference of light-excitations.
It should be noted that the steady-state coherence or, equivalently, efficiency is
the combined result of light-induced transient coherence, inhomogeneous depletion, and system-bath correlation,
and is thus not necessarily correlated with quantum beatings.
\end{abstract}
\maketitle

\newpage

Ever since the first evidence for quantum coherence was demonstrated in photosynthetic systems,
the role of quantum coherence in the light-harvesting process has inspired numerous scientific studies.\cite{Scholes2017,Cao2020}
In particular, questions such as whether quantum coherence can be initiated by incoherent sunlight and whether coherence
plays a role in the function of light-harvesting complexes have been discussed extensively in literature.
\cite{Cheng2009,Mancal2010,Ishizaki2012,Turner2013,Olsina2014,Tscherbul2014,Brumer2018,Shatokhin2018,Chan2018}
Previous calculations\cite{Olsina2014} have shown that, for the parameters relevant to photosynthetic systems, the exciton dynamics initiated by incoherent light exhibits dynamical coherence (quantum beatings) on the sub-picosecond
timescale; however, the transient coherent time-scale may not be sufficiently long for the beatings to play a crucial role in efficient energy
transfer to the reaction centers.  Yet, in natural systems,  it is the non-equilibrium steady state (NESS)
of the light-harvesting process that determines their functions, thus motivating the steady-state analysis reported here.

In addition to experimental relevance, our theoretical analysis is also inspired by
previous studies of steady-state coherence in specific configurations of model systems. \cite{Kozlov2006,Xu2016B,Tscherbul2018}
In particular, it has been shown that non-vanishing steady-state coherence can enhance the efficiency of photosynthetic units and photovoltaic devices.\cite{Scully2011,Dorfman2013,Xu2016A,Dorfman2018,Zhang2015,Rouse2019}
In spite of these results, for molecular systems weakly driven by incoherent light,
a general analytic theory is still lacking.  More importantly,  there is an urgent need for the community to
elucidate  how steady-state  coherence
relates to detailed balance, energy transfer flux, optimal efficiency, and choice of basis set.
In this work, we first address these open questions quantitatively using a light-driven dimer model and then extend to general quantum networks
to reveal the crucial role of  steady-state coherence in light-harvesting  energy transfer.

\begin{figure}
\centerline{\scalebox{0.3}{\includegraphics[trim=20 0 0 0,clip]{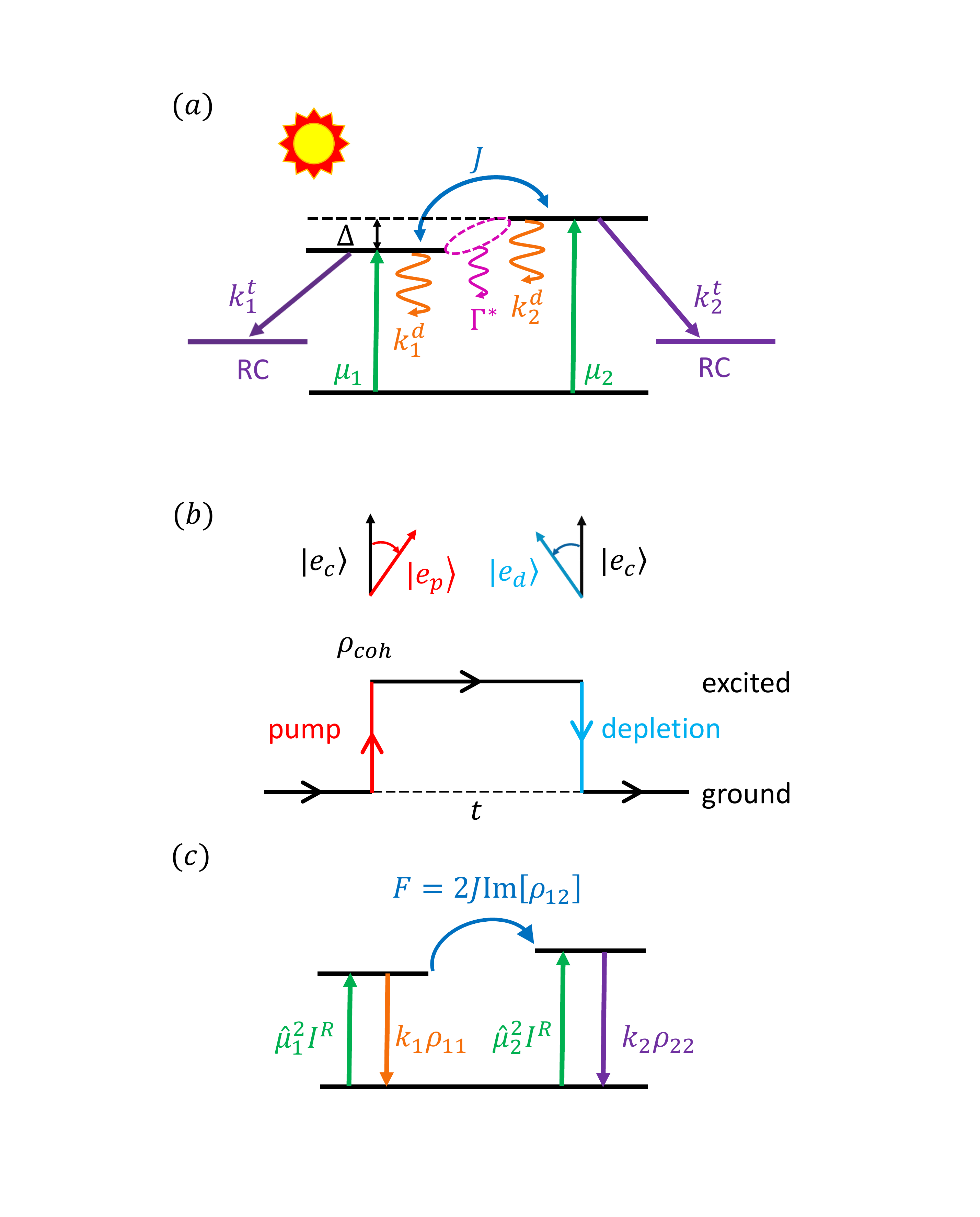}}}
\caption{(a) Schematic of the dimer system consisting of two excited states with detuning $\Delta$ and a common ground state. The excited states are coupled with intermolecular coupling $J$. The depletion of excited population of each molecule is quantified by $k_i=k_i^t+k_i^d$, where $k_i^t$ characterizes the trapping to the reaction center and $k_i^d$ characterizes the non-radiative irreversible decay to the ground state. Each molecule is further coupled to local environment leading to dephasing at a rate of $\Gamma^*$. Non-equilibrium dynamics of the dimer system is driven by the interactions between the transition dipole moments of the dimer $\boldsymbol{\mu}_i$ and incoherent sunlight. (b) The schematic diagram for the rotations of the eigenstates  by incoherent excitation (red line) and population depletion (blue line),
where $|e_c\rangle$, $|e_p\rangle$, and $|e_d\rangle$ denote respectively the exciton states, and eigenstates during the excitation
and depletion processes. Note the rotation of the eigenstates is proportional to $\boldsymbol{\mu}_+\cdot\boldsymbol{\mu}_-=(\mu_1^2-\mu_2^2)/2$ in the excitation process and proportional to $-\delta k$ in the depletion process. (c) The schematic of steady-state population flux in the dimer system.}
\label{Fig1}
\end{figure}

To begin, we consider a generic molecular dimer illustrated in Fig.~\ref{Fig1} (a), which captures the essential physics relevant for the light-harvesting process. In particular, we adopt delocalized photon excitation and localized depletion,
but do not include delocalized radiative decay as it occurs on much slower time scale.\cite{Cao2009}
The case of delocalized trapping is analyzed in S5 of the Supporting Information and summarized later in the context of 'general dimer model'.
The dynamics of the molecular dimer is dictated by the quantum master equation,
\begin{align}\label{EOM}
\dot{\rho}(t)=-[L_{sys}+L_{trap}+L_{decay}+L_{dissip}]\rho (t)+\rho^0,
\end{align}
where the reduced Planck constant is set to unit $\hbar =1$ hereafter.
Here, the Liouville superoperator $[L_{sys}]_{ij,kl}=i(H_{ik}\delta_{jl}-H_{lj}\delta_{ik})$ with $H_{ij}=(1-\delta_{ij})J+\delta_{ij}\varepsilon_i$,
where $J$ is the intermolecular coupling and $\varepsilon_i$ is the site energy of molecule $i$.
In the paper, we adopt the excitonic Hamiltonian and excitonic coherence, but the same treatment can apply
to vibronic states and any other molecular states as well.\cite{beatings}
The population depletion of the dimer system originating from local
energy trapping at the reaction center and irreversible decay to the ground state at each molecule are characterized by the Liouville superoperators $[L_{trap}+L_{decay}]_{ij,ij}=(k_i+k_j)/2$ with $k_i=k_i^t+k_i^d$, where $k_i^t$ and $k_i^d$ are phenomenological
trapping and decay rates at molecule $i$, respectively. The dissipation of the dimer due to the coupling to the environment
is considered as pure dephasing in the Haken-Strobl-Reineker (HSR) model \cite{Haken1972}, where $[L_{dissip}]_{ij,ij}=(1-\delta_{ij})\Gamma^*$ with the pure dephasing rate $\Gamma^*$.
Based on the white noise approximation,\cite{Olsina2014} we have shown that
the stationary incoherent sunlight induces a pure state given by
\begin{equation}
\rho^0 = [\rho^0_{11}, \rho^0_{12}, \rho^0_{21}, \rho^0_{22}] =
I^R[\hat{\mu}_1^2, \hat{\boldsymbol{\mu}}_1\cdot\hat{\boldsymbol{\mu}}_2, \hat{\boldsymbol{\mu}}_1\cdot\hat{\boldsymbol{\mu}}_2, \hat{\mu}_2^2],
\label{rho0}
\end{equation}
where $\hat{\boldsymbol{\mu}}_i=\boldsymbol{\mu}_i/\bar{\mu}$ is the normalized transition dipole,  $\boldsymbol{\mu}_i$ is the transition dipole moment of molecule $i$,  and $\bar{\mu}=\sqrt{\mu_1^2+\mu_2^2}$ is the magnitude of the total dipole moment. The master equation (\ref{EOM}) shows the generic interplay between the incoherent excitations and the population depletion and can reduce to the special cases discussed in the literature (see Sec.~S1B in Supporting Information).

\textit{Excitonic coherence: Detailed balance and decomposition.}
The steady-state solution to Eq.~(\ref{EOM}) is derived and analyzed in Sec.~S1 of Supporting Information. For simplicity of presentation,
we first consider the special case of the degenerate dimer ($\varepsilon_1=\varepsilon_2=\varepsilon$) without environmental effects ($\Gamma^*=0$), and then extend our conclusions to the general dimer model.
The non-equilibrium steady-state (NESS) in the exciton basis is solved in Supporting Information, giving
\begin{subequations}
\label{rho_steady}
\begin{align}
\rho_{++}&=\frac{\hat{\mu}_+^2I^R}{\bar{k}}+\frac{\delta k}{2J}{\rm Im}[\rho_{+-}], \\
\rho_{--}&=\frac{\hat{\mu}_-^2I^R}{\bar{k}}+\frac{\delta k}{2J}{\rm Im}[\rho_{+-}], \\
\rho_{+-}&=-\frac{A I^R}{2}\frac{1-i2J/\bar{k}}{(2J)^2+k_1 k_2},
\end{align}
\end{subequations}
where $\bar{k}=(k_1+k_2)/2$ and $\delta k=(k_1-k_2)/2$ and $\hat{\boldsymbol{\mu}}_{\pm}=(\hat{\boldsymbol{\mu}}_1\pm\hat{\boldsymbol{\mu}}_2)/\sqrt{2}$ are defined in the exciton basis.
A key prediction of the steady-state solution is the relationship between the exciton populations and coherence.
Specifically, the first terms in Eq.~(\ref{rho_steady}a) and Eq.~(\ref{rho_steady}b) are the local contributions,
which are the steady state for each exciton level without coherent mixing and are determined by the
balance between the corresponding pumping and depletion rate.
The second terms are the coherent mixing contributions, which are identical for both exciton states and
are proportional to the imaginary part of quantum coherence ${\rm Im}[\rho_{+-}]$.
Further, the magnitude of the steady-state exciton coherence is proportional to $A\equiv\hat{\mu}_2^2 k_1-\hat{\mu}_1^2 k_2$.
The parameter $A$ measures the deviation from detailed balance,
\begin{align}
\label{DB}
\frac{\hat{\mu}_1^2}{k_1}=\frac{\hat{\mu}_2^2}{k_2},
\end{align}
indicating a constant ratio of the excitation and depletion rates in the dimer.
In some early analysis,\cite{Tscherbul2014} the light-harvesting system is coupled to a single thermal light source (e.g. blackbody radiation),
where detailed balance is automatically observed. Then, coherence vanishes as the system relaxes to the thermal equilibrium.
In order to break detailed balance and induce steady-state coherence,
light-harvesting systems must couple to at least another thermal bath, such as the protein environment or the reaction center, in
addition to sunlight radiation.\cite{Olsina2014,Xu2016A,Tscherbul2018,Koyu2020}

\begin{figure}
\centerline{\scalebox{0.34}{\includegraphics[trim=0 0 0 0,clip]{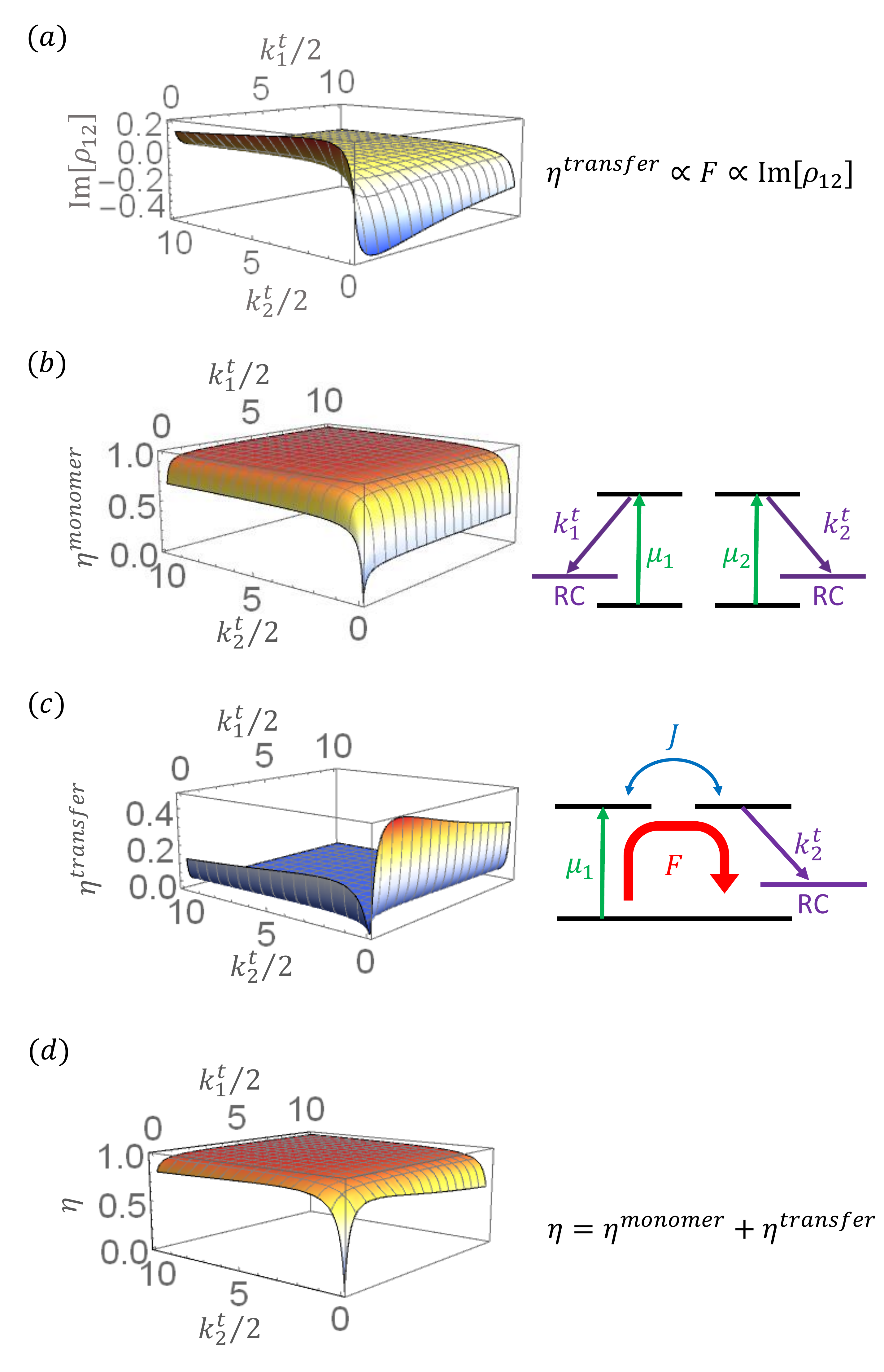}}}
\caption{Quantum coherence and energy transfer efficiency as a function of trapping rates $k_i^t$, with $J=1$, $k_d=0.25$, and $|\boldsymbol{\mu}_1|/|\boldsymbol{\mu}_2|=2/3$: (a) ${\rm Im [\rho_{12}]}$ in the unit of $I^R$; (b) monomer contribution $\eta^{monomer}$; (c) transfer contribution $\eta^{transfer}$; and (d) $\eta=\eta^{monomer}+\eta^{transfer}$.}
\label{Fig2}
\end{figure}

An equilibrium system weakly perturbed by thermal noise assumes the Boltzmann distribution in its eigenstates with zero exciton coherence.  Therefore, the non-vanishing steady-state exciton coherence arises from non-equilibrium driving, i.e, the excitation by incoherent sunlight and the depletion from the excitation manifold. To quantify these two contributions, we decompose the NESS coherence in Eq.~(\ref{rho_steady}c)
into (see Sec.~S2 in Supporting Information),
\begin{align}
\rho_{+-}=&\rho^{light}_{+-}+\rho^{depletion}_{+-}\notag\\=&
\bigg[\bar{k}\rho^0_{+-}-{ \delta k \over 2} \big(\rho^0_{++}+\rho^{0}_{--}\big)\bigg]
\frac{1-i2J/\bar{k}}{(2J)^2+k_1k_2},
\label{coherence}
\end{align}
where
$\rho^{0}_{\pm\pm}=\hat{\mu}^2_\pm I^R$, $\rho^{0}_{+-}
=(\hat{\boldsymbol{\mu}}_+\cdot\hat{\boldsymbol{\mu}}_-)I^R$
are the light-induced initial state $\rho^0$ in Eq.~(\ref{rho0}) in the exciton basis.
Thus, the steady-state coherence can be induced by excitation and depletion,
represented by the first and second terms in the brackets, respectively.
Specifically, during the evolution of the light-harvesting dimer, its basis set rotates twice:
The first rotation arises from the excitation process, where the incoherent radiation field creates the initial
coherence $\rho^{0}_{+-}$, characterized by the effective  transition dipole
$\hat{\boldsymbol{\mu}}_+\cdot\hat{\boldsymbol{\mu}}_-=(\hat{\mu}_1^2-\hat{\mu}_2^2)/2$, which is related
to the quantum beatings discussed previously.\cite{Olsina2014}
The second rotation occurs when the dimer system experiences asymmetric depletion  ($\delta k\neq 0$)
and is proportional to the initial population $\rho^0_{++}+\rho^0_{--}$.
As shown in Fig.~\ref{Fig1}(b), the superposition of the two rotations is constructive when they are in phase, i.e.,
$\delta\mu\delta k<0$ ($\delta \mu=\mu_1-\mu_2$), and is destructive when they are out of phase, $\delta\mu\delta k > 0$.
An interesting finding from the previous analysis\cite{Olsina2014} is  that light-induced transient coherence, as manifested as quantum beatings,
relaxes on a time-scale faster than the typical energy transfer timescale. However, Eq.~(\ref{coherence}) suggests that the exciton
population can also contribute due to the asymmetry in population depletion and thus the light-induced
beatings may not dominate the steady-state coherence in photosynthetic systems.\cite{Olsina2014}

\textit{Intermolecular coherence: Flux and conservation laws.}
To further explore the role of quantum coherence in light-harvesting energy transfer,
we transform the steady-state solution in Eq.~(\ref{rho_steady}) to the local site basis (i.e. molecular basis), giving
\begin{subequations}\label{rho_steady_site}
\begin{align}
\rho_{11}=&\frac{\hat{\mu}_1^2I^R}{k_1}- {2J \over k_1} {\rm Im}  [\rho_{12}]
 =  {1\over k_1} (\rho^{0}_{11} - F ), \\
\rho_{22}=&\frac{\hat{\mu}_2^2I^R}{k_2}+ {2J \over k_2} {\rm Im}  [\rho_{12}]
= {1\over k_2}  ( \rho^{0}_{22} + F ), \\
\rho_{12}=&\frac{I^R}{\bar{k}}\Big(\hat{\boldsymbol{\mu}}_1\cdot\hat{\boldsymbol{\mu}}_2-i\frac{AJ}{4J^2+k_1k_2}\Big).
\end{align}
\end{subequations}
In Eq.~(\ref{rho_steady_site}a) and Eq.~(\ref{rho_steady_site}b),
the first terms are the incoherent monomer contributions,
arising from the kinetic balance of local excitation and local population depletion in each molecule. The second terms are the coherent transfer contributions with opposite signs, indicating that the steady-state coherence induces a transfer flux
between the two molecules of the dimer
(see Fig.~\ref{Fig1}(c) and Sec.~S1A in Supporting Information), i.e.,
\begin{align}
F=2J{\rm Im}[\rho_{12}] \propto A.
\label{flux}
\end{align}
The coherence-flux relation is completely general and was introduced\cite{Wu2012} earlier in the context of light-harvesting energy transfer and applied to FMO and other model systems.
Because of the local conservation of exciton population,\cite{Wu2012}
the transfer flux at any molecular site in a light-harvesting network is summed to zero.
The second equality in Eq.~(\ref{flux}) demonstrates that the deviation from detailed balance,  characterized by $A$,
drives the non-equilibrium transfer between the two molecules via quantum coherence.
Further, we note
\begin{align}
{\rm Im}[\rho_{12}]=-{\rm Im} [\rho_{+-}],
\end{align}
such that the imaginary part of coherence
is invariant to the choice of the basis set, which is an interesting observation given the recent discussion
about the correct basis set to define quantum coherence in light-harvesting energy transfer.

In Fig.~\ref{Fig2} (a), we plot the NESS transfer flux as a function of the trapping rates $k^t_1$ and $k^t_2$,
assuming the irreversible decay rates are the same for the dimer ($k_1^d=k_2^d=k^d$).
The transfer flux is significant when either of the trapping rates is small and comparable to the decay rate,  $k_i^t\approx k^d$,
but vanishes when both trapping rates are large, i.e., $k^t_1 \gg k^d$ and $k^t_2 \gg k^d$.
When $k_1^t$ ($k_2^t$) is smaller, the excitation flux flows from molecule $1$ ($2$) to $2$ ($1$), the sign of flux is positive (negative).
Note that the magnitude of the flux is proportional to $\hat{\mu}_i^2$, which explains the asymmetry
of the diagram with $|\boldsymbol{\mu}_1|<|\boldsymbol{\mu}_2|$ in Fig.~\ref{Fig2}.  Later, it will be shown that
the coherent part of efficiency $\eta^{transfer} $ is proportional to $F$, thus explaining the similarity between Fig.~\ref{Fig2} (a) and
Fig.~\ref{Fig2}(c).

From the solution in Eq.~(\ref{rho_steady_site}), we obtain the NESS population flux in the dimer system, yielding
\begin{equation}
I^R = k_1\rho_{11}+k_2\rho_{22},
\label{population}
\end{equation}
where $I^R$ describes excitation by incoherent light and the right-hand side describes population depletion.
As expected, these two fluxes are equal as the exciton population is conserved in the steady-state limit.
Interestingly, the  excitation rate, $I^R$, is determined by the light intensity and the magnitudes of transition dipoles, but is independent of molecular configurations in light-harvesting complexes, such as
the dipole orientation, intermolecular distance, or dipole-dipole interaction,

In the classical description,
each molecule carries a fixed amount of excitation energy so the energy flux is simply the product of the excitation energy
and exciton population flux,
$\varepsilon I^R$.  This picture is modified in the presence of quantum coherence because the excitation energy is delocalized.\cite{Xu2016A}
To see this quantitatively, we derive the excitation energy flux at the steady state  (see Sec.~S3 in Supporting Information), giving
\begin{equation}
(\varepsilon+2J\hat{\boldsymbol{\mu}}_1\cdot\hat{\boldsymbol{\mu}}_2)I^R=\varepsilon(k_1\rho_{11}+k_2\rho_{22})+2J\bar{k}{\rm Re}[\rho_{12}].
\label{energy}
\end{equation}
where the left and right hand sides of the equation correspond to energy excitation and depletion, respectively,
and are equal because of energy conservation.
The first terms are exactly the classical result, $\varepsilon I^R= \varepsilon (k_1\rho_{11}+ k_2\rho_{22})$,
whereas the second terms are the quantum corrections,
$2J I^R \hat{\boldsymbol{\mu}}_1\cdot\hat{\boldsymbol{\mu}}_2=2J\bar{k}{\rm Re}[\rho_{12}]$.
Unlike the population flux, the quantum correction depends on the molecular configuration and
is proportional to the real part of coherence,  ${\rm Re}[\rho_{12}]$.
Thus, the real and imaginary parts of quantum coherence have clear but different physical meanings in energy transfer.

\textit{Optimal efficiency.}
The excitation energy in light-harvesting systems can be trapped at the reaction center with rate of $k^t$ or dissipated via radiative
or non-radiative channels with rate of $k^d$.
Then, the efficiency of light-harvesting energy transfer $\eta$ can be defined
as the trapping probability \cite{Cao2009,Rebentrost2009,Chin2010,Wu2010}, giving
\begin{align}\label{q_def}
\eta=\frac{\sum_ik_i^t\rho_{ii}}{\sum_ik_i^t\rho_{ii}+\sum_ik_i^d\rho_{ii}}={1\over I^R} \sum_ik_i^t\rho_{ii},
\end{align}
where the denominator is total exciton flux in Eq.~(\ref{population}), $\sum_ik_i^t\rho_{ii}+\sum_ik_i^d\rho_{ii}=I^R$.
Inserting the steady-state density matrix in Eq.~(\ref{rho_steady_site}) into Eq.~(\ref{q_def}), we obtain,
\begin{align}\label{q}
\eta=&\eta^{monomer}+\eta^{transfer} \notag\\
=&\sum_{i=1}^2\kappa_i\hat{\mu}_i^2-\frac{F}{I^R}(\kappa_1-\kappa_2),
\end{align}
which consists of monomer and transfer contributions. Here, $\kappa_i = k_i^t/k_i$ is the branching ratio for trapping at the $i$-th site. For the monomer contribution, excitation and trapping occur locally at the same site, as shown schematically in Fig.~\ref{Fig2}(b),
and the efficiency is the sum of local trapping probabilities.
For the transfer contribution, the excitation is pumped at one site and transfers to the other site, as shown schematically in Fig.~\ref{Fig2}(c).
In light-harvesting systems, light-absorption and trapping usually occur on different molecules, so the efficiency is dominated
by the transfer part, giving
\begin{align}
\label{relation}
\eta \propto F \propto {\rm Im } [\rho_{12}],
\end{align}
Thus, efficiency $\eta$ is proportional to exciton transfer flux, $F$,
which in turn is determined by quantum coherence ${\rm Im} [\rho_{12}]$.  Derived explicitly for the light-harvesting dimer,
the coherence-flux-efficiency relation will be established later in Eq.~(\ref{network}) for arbitrary quantum networks.

\begin{figure}
\centerline{\scalebox{0.34}{\includegraphics[trim=50 50 0 0,clip]{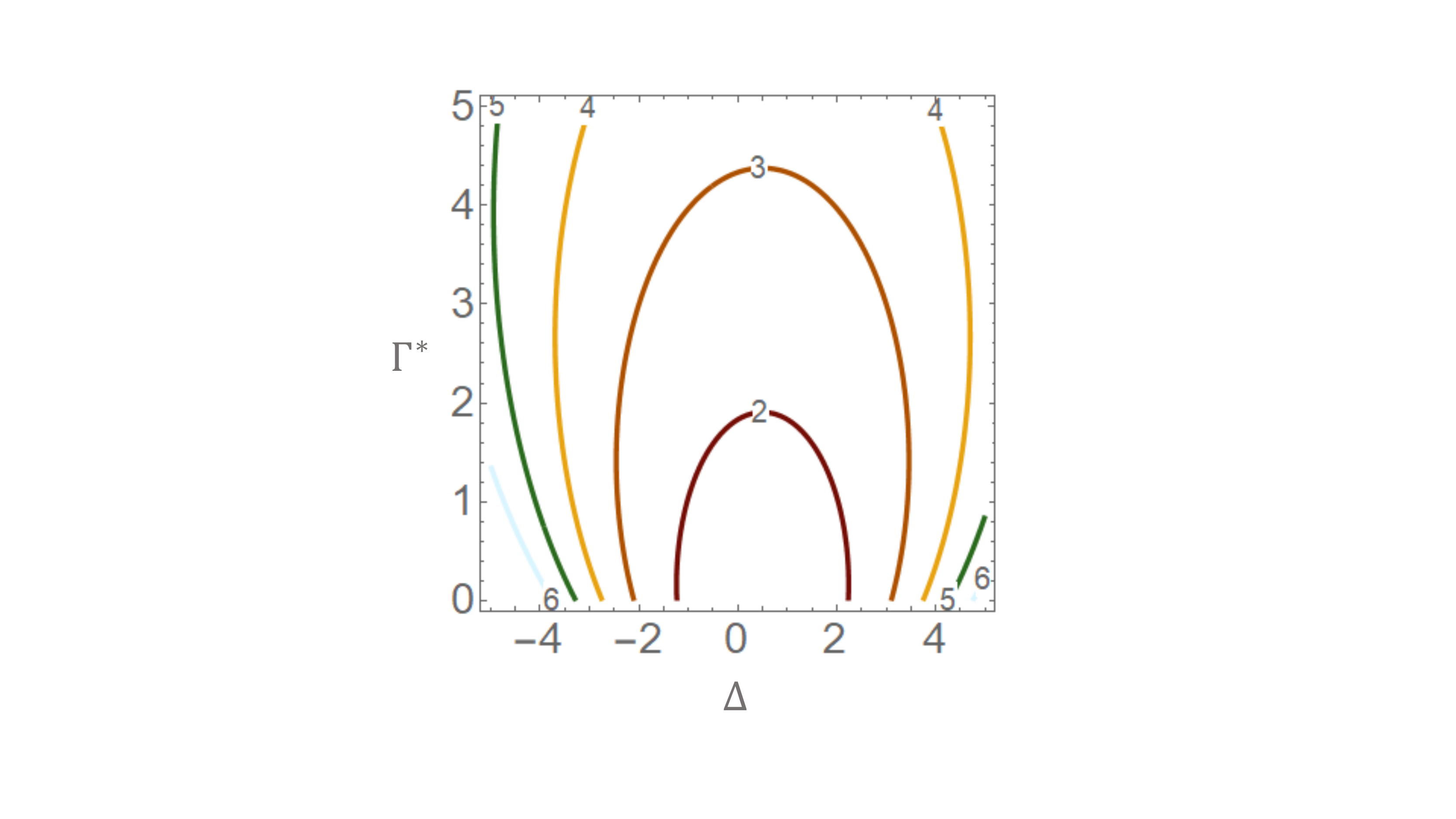}}}
\caption{The contour of average trapping time as a function of detuning $\Delta=\varepsilon_2-\varepsilon_1$ and dephasing rate $\Gamma^*$.
The model parameters are $J=1$, $k_2^t=3$, $|\boldsymbol{\mu}_1|/|\boldsymbol{\mu}_2|=2$,
and $\boldsymbol{\mu}_1//\boldsymbol{\mu}_2$.}
\label{Fig3}
\end{figure}

In Fig.~\ref{Fig3}, we plot $\eta^{monomer}$ and $\eta^{transfer}$ as a function of the trapping rates and can clearly identify two regimes.
In the first regime (see Fig.~\ref{Fig2}(b)), where both trapping rates are large, i.e.,  $k_i^t\gg k^d$,
efficiency is dominated by the monomer contribution $\eta^{monomer}$ as light absorption and energy trapping occur at the same molecule.
When the two trapping rates are taken to be identical,  $k_1^t = k_2^t\equiv k^t$, the energy transfer efficiency reduces to,
\begin{align}
\eta\simeq  \frac{1}{1+k^d \langle t\rangle } = \frac{1}{1+k^d/k^t},
\end{align}
where the first equality is a general relation\cite{Cao2009} that approximates the efficiency with the average trapping time
$\langle t\rangle$ and the second equality gives  $\langle t\rangle=1/k^t$ for this special case.
Evidently, efficiency approaches $1$ as $k^t$ approaches infinity,
so there is no non-trivial optimization for local transfer in the monomer regime.

In the second regime (see Fig.~\ref{Fig2} (c)), where either of the trapping rates is small, i.e.,  $k_i^t\le k^d$,
efficiency is dominated by the transfer contribution $\eta^{transfer}$  and
excitation energy absorbed at one molecule is transferred to the reaction center at the other molecule.
As observed in Fig.~\ref{Fig2}(c),  in the transfer regime,
there are apparent non-trivial optimal trapping rates for the maximal
efficiency.\cite{Cao2009,Wu2013PRL,Jesenko2013,Leon2014,Zhang2017,Zerah-Harush2018,Dutta2020,Tomasi2020}
For example, with $k_1^t\approx k^d$,
the average trapping time is given as  (see Sec.~S4 in Supporting Information),
\begin{align}\label{t_2}
\langle t \rangle = \frac{2}{k_2^t}+\hat{\mu}_1^2\frac{k_2^t}{4J^2}.
\end{align}
which has a minimal as a function of $k_2^t$.  In Eq.~(\ref{t_2}), $\hat{\mu}_1^2$ denotes the fraction
of the delocalized excitation in the dimer.
When $\hat{\mu}_1^2=1$, the trapping time reduces to Eq.~(4) in Ref.~\citenum{Cao2009}, an early result.
Fig.~\ref{Fig2} (d) plots the sum of the two contributions. Interestingly, the monomer and transfer regimes are complementary so that
the overall efficiency remains high over the entire parameter space, except in the small regime where both trapping rates are small.

In the above, we have adopted a definition of energy transfer efficiency based on the exciton population flux; yet,
the analysis remains valid even when energy flux is used  instead of population flux.
Based on Eq.~(\ref{energy}), we define efficiency in terms of the energy flow to the reaction center as
\begin{align}\label{JRC}
\eta_{\epsilon}=  { \varepsilon \eta + 2J\hat{\boldsymbol{\mu}}_1\cdot\hat{\boldsymbol{\mu}}_2 \bar{\kappa}
\over \varepsilon+ 2J\hat{\boldsymbol{\mu}}_1\cdot\hat{\boldsymbol{\mu}}_2 }
\end{align}
where $\bar{\kappa}=(\kappa_1+\kappa_2)/2$.
The first term in the numerator is exactly the site energy $\varepsilon$ multiplied by population transfer efficiency $\eta$ defined in Eq.~(\ref{q_def}), which is the prediction of the classical picture, and the second term is the quantum correction,
which is proportional to $\hat{\boldsymbol{\mu}}_1\cdot\hat{\boldsymbol{\mu}}_2$. In light-harvesting systems, the site energy $\varepsilon$ is much larger than the excitonic couplings, $\varepsilon \gg J$, so the second term is negligible and we have $\eta_\epsilon \simeq \eta$.\cite{Manzano2013}


\textit{General dimer model.}
Though we have focused on the degenerate dimer model for simplicity, the solution presented in Supporting Information is for the general dimer model and the above analysis remains valid. Specifically, when the detuning $\varepsilon_2-\varepsilon_1=\Delta$ and dephasing $\Gamma^*$ are considered,
the populations of the dimer remain the same as in Eq.~(\ref{rho_steady_site}),
whereas the expression for quantum coherence is modified, giving
\begin{align}
{\rm Im}[\rho_{12}]=-{\rm Im}[\rho_{+-}]=-I^R\frac{JA\Gamma-\Delta(\hat{\boldsymbol{\mu}}_1\cdot\hat{\boldsymbol{\mu}}_2)k_1k_2}{4J^2\bar{k}\Gamma+(\Delta^2+\Gamma^2)k_1k_2},
\end{align}
where $\Gamma=\Gamma^*+\bar{k}/2$.
Clearly, in the presence of detuning $\Delta$,  quantum coherence also can be created by the interference of the transition dipole moments $\boldsymbol{\mu}_1\cdot\boldsymbol{\mu}_2$ such that the detailed balance relation needs to be supplemented.
Apart from this difference, our previous predictions
including transfer flux, basis set invariance, decomposition, and flux conservation relations remain valid.

As $\Delta$ and $\Gamma^*$  appear only in the transfer contribution $\eta^{transfer}$, we examine the energy transfer efficiency in the transfer regime.
For convenience of analysis, we assume $k_1^d=k_2^d=k^d$, $k_1^t < k^d$, and $k^d/J\ll 1$, and obtain the average trapping time as,
\begin{align}\label{t_2_diss}
\langle t \rangle= \frac{2}{k_2^t}+\hat{\mu}_1^2\frac{1}{2J^2}\frac{\Gamma^2+\Delta^2}{\Gamma}-\hat{\boldsymbol{\mu}}_1\cdot\hat{\boldsymbol{\mu}}_2\frac{\Delta}{J\Gamma}.
\end{align}
where the last two terms arise from quantum coherence.
Eq.~(\ref{t_2_diss}) reduces to Eq.~(4) in Ref.~\citenum{Cao2009} when $\mu_2=0$, which exhibits non-trivial optimal dephasing rate
in the non-degenerate dimer ($\Delta\neq 0$).  In the general case,
in addition to the dephasing rate, the detuning and the transition dipoles also have non-trivial optimal conditions because of the
interference as given in the last term of Eq.~(\ref{t_2_diss}). In Fig.~\ref{Fig3}, the average trapping time Eq.~(\ref{t_2_diss})
is examined as a function of $\Delta$ and $\Gamma^*$, which shows a global optimal slightly away from the degenerate condition.

Before moving onto quantum networks, we briefly discuss the case of delocalized trapping by the introduction of the delocalized depletion rate $k_{12}$.  Then,  the total exciton flux becomes
\begin{align}
I^R=k_1\rho_{11}+k_2\rho_{22}+2k_{12}{\rm Re}[\rho_{12}].
\end{align}
where the exciton population created by incoherent excitation $I^R$
decays through both localized and delocalized depletions. Following the steady-state solution in S5 of Supporting Information, we obtain
 the formal expression for efficiency
\begin{align}
\eta=&\eta^{monomer}+\eta^{transfer}+\eta^{delocalized}\notag\\
=&\sum_{i=1,2} \kappa_i \hat{\mu}_i^2
+\frac{1}{I^R}(\kappa_2-\kappa_1) F
+ \frac{1}{I^R}(2- \kappa_1 -\kappa_2)k_{12}^t{\rm Re}[\rho_{12}],
\end{align}
which can be decomposed into monomer, transfer, and delocalized trapping contributions, Evidently, efficiency is correlated to both the real and imaginary parts of quantum coherence but with different physical meanings,
consistent with a recent calculation reported in Ref.~\citenum{Jung2020}.
The delocalized trapping is a simplified description of generalized F\"{o}rster energy transfer and super-radiance. A quantitative
description of these collective processes requires the consideration of the system-bath correlation,\cite{Ma2015} which is beyond the scope of
this paper.

\begin{figure}
\centerline{\scalebox{0.34}{\includegraphics[trim=50 50 0 0,clip]{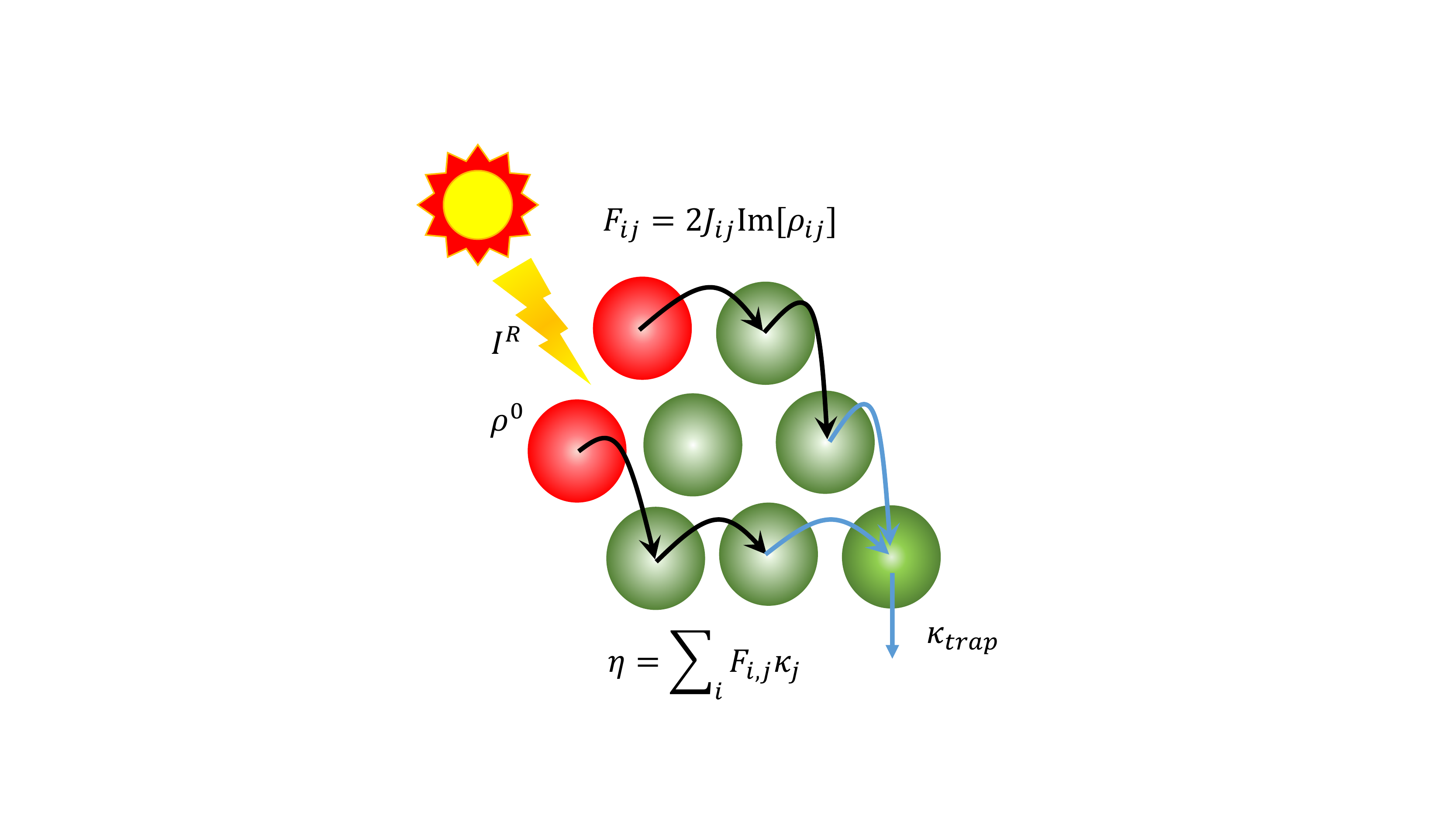}}}
\caption{Schematic diagram of a light-harvesting network. The molecules in red color represent the chromophores excited by incoherent sunlight
and the molecule in dark green represents the chromophore directly connected to the reaction center.   The incoherent sunlight with intensity of $I^R$ creates a stationary initial state $\rho^0$ under the white noise approximation.   The excitation energy transfers through NESS flux between molecules $F_{ij}=2J_{ij}{\rm Im}[\rho_{ij}]$ and finally traps at the reaction center  with branching ratio $\kappa_{trap}$.
Then, the energy transfer efficiency can be expressed as $\eta = c \sum_i F_{i,j}\kappa_j = c \sum_i F_{i,trap}\kappa_{trap}$, where $c=1/I^R$ is the normalization constant.}
\label{Fig4}
\end{figure}

\textit{Light-harvesting networks.}
It is straightforward to extend the master equation (\ref{EOM}) to an arbitrary quantum network, (see Fig.~\ref{Fig4})
i.e, a multi-chromophoric system or multi-level exciton system, with the excitonic coupling $J\rightarrow J_{ij}$ and the dephasing rate $\Gamma^*\rightarrow\Gamma^*_{ij}$.
In the steady-state limit, the exciton population at site $i$ is given in the form of (see S6 of Supporting Information)
\begin{align}
\rho_{ii} =  \frac{1}{k_i} ( \hat{\mu}_i^2I^R - \sum_{j\neq i}F_{ij} ),
\end{align}
where the first term is the incoherent local contribution from monomers and the second term is the exciton transfer contribution characterized by the NESS flux between a pair of molecules,
\begin{align}
F_{ij}=2 J_{ij}{\rm Im} [\rho_{ij}].
\end{align}
As shown earlier in Ref.~\citenum{Wu2013},
the flux thus defined  characterizes the energy transfer pathways in a non-equilibrium quantum network (also see Fig.~\ref{Fig4}).
Since the population flux is in the same form as in Eq.~(\ref{population}),
$I^R=\sum_i k_i \rho_{ii}$, the energy transfer efficiency is then given by,
\begin{align}
\eta=\sum_i\kappa_i\hat{\mu}_i^2-\frac{1}{I^R}\sum_{i<j}(\kappa_i-\kappa_j)F_{ij}
= \sum_i\kappa_i\hat{\mu}_i^2 + \frac{1}{I^R} \sum_{i \neq j}F_{ij} \kappa_j
\end{align}
where the identity $F_{ij}=-F_{ji}$ is used to arrive the last expression.
This definition has an intuitive interpretation based on network kinetics:  The first term is the branching probability resulting from local excitation and depletion associated with monomers, whereas the second term is the sum of all trapping flux to the reaction center.  Here, trapping flux is understood
as the product of transfer flux F and branching ratio $\kappa$.  As in Eq.~(\ref{relation}) for dimers, we now formally establish the
coherence-flux-efficiency relation,
\begin{align}
\label{network}
\eta = \frac{1}{I^R} \sum_{i \neq j}F_{ij} \kappa_j = \frac{2}{I^R} \sum_{i \neq j} J_{ij} {\rm Im} [{\rho}_{ij}] \kappa_j
\end{align}
where we assume that the excitation and trapping occur on difference molecules.
Eq~(\ref{network}) holds generally for any exciton networks of arbitrary connectivity
under the stationary condition, which is not limited to incoherent radiation or incoherent pumping.

\textit{Discussion.}
Light-harvesting systems are composed of many pigments, proteins, reaction centers, etc, and are thus far more complex than the model systems
studied here.
Yet, our basic coherence-flux-efficiency relation remains valid. For applications to light-harvesting systems, we now comment  on realistic considerations:
\begin{enumerate}
\item
For simplicity, the environment is treated classically as white noise, which yields dephasing but not thermalization.
Our previous studies via numerical simulations and polaron calculations have shown that a quantum thermal bath leads
to stationary coherence as a result of the non-canonical thermal distribution due to system-bath correlation.~\cite{Olsina2014,Ma2015}
Thus, the steady-state coherence is
the combined result of light-induced beatings, inhomogeneous depletion, and thermalization, and is thus not necessarily
correlated  with quantum beatings in 2D spectra.\cite{Engel2007,Collini2010}  Therefore, efficient energy transfer does not necessarily require
long-lived quantum beatings (i.e., light-induced transient coherence),
which is consistent with recent experimental evidence in FMO.\cite{Duan2017,Thyrhaug2018}
\item
For the dimer model, quantum coherence defined by the imaginary part of the density matrix is basis-set invariant.
Beyond the dimer model,  quantum coherence depends on the choice of basis set, thus raising the question of the right basis.\cite{Tomasi2020}
Yet, physical measurements are independent of the choice of basis,
so this question is for the convenience of theoretical description and numerical approximation.
Typically, the local basis is a natural choice for describing transport, whereas the exciton basis is more convenient
for calculating spectroscopy.  Since light-harvesting systems are strongly coupled to the protein environment,
the polaron basis provides a physically-motivated description.\cite{Xu2016B}
\item
The flux-coherence relation can be defined between any pairs of molecules, $F_{ij}=J_{ij}{\rm Im}[\rho_{ij}]$,
and is shown to be general in Appendix B of Ref.~\citenum{Wu2012}.
Interestingly,  this relation reduces to the classical flux in the strong damping limit,
$F_{ij}=\rho_i k_{ij} -\rho_j k_{ji}$,  where $k$ is the hopping rate (e.g., F\"{o}rster rate for energy transfer).
This classical limit is the leading term of a systematic mapping of quantum networks to kinetic networks,\cite{Wu2013}
which allows a perturbative calculation of long-range transfer.\cite{Skourtis1995}
In FMO, the parametric dependence of
energy transfer efficiency can be reproduced by the classical flux, suggesting the dominance of the hopping mechanism.\cite{Wu2012}
 Yet, regardless of step-wise hopping or wave-like propagation, the coherence-flux-efficiency relation holds rigorously and generally.
 \item
In photosynthetic systems, the number of light-absorption pigments are larger than the number of reaction centers, so that excitation energy
funnels to the reaction centers driven by energetic and entropic gradients.\cite{Moix2011,Adolphs2006}
In this case, light-harvesting systems can be optimized for their functions, not only via coherence
but also via composition and spatial arrangements. The optimization of self-assembly superstructures
has been studied in purple bacteria membranes as an illustrative example of light-harvesting networks.\cite{Hu1998,Cleary2013}
\end{enumerate}

In summary, we have demonstrated that steady-state coherence leads to optimal energy transfer
in light-harvesting systems.
Specifically, as given explicitly in Eq~(\ref{network}), efficiency $\eta$ is proportional to exciton transfer flux, $F$,
which in turn is determined by quantum coherence ${\rm Im} [{\rho}]$.
The coherence-flux-efficiency relation holds rigorously and generally for any exciton networks of arbitrary connectivity
under the stationary condition, which is not limited to incoherent radiation or incoherent pumping.
For light-harvesting networks under incoherent sunlight,
non-equilibrium energy transfer flux is
driven by the breakdown of detailed balance and by quantum interference of light-excitations.
It should be noted that the steady-state coherence or, equivalently, the energy transfer flux is
the combined result of light-induced transient coherence, inhomogeneous depletion, and system-bath correlation,
and is thus not necessarily correlated with quantum beatings in 2D spectra.
These findings reveal the crucial role of steady-state quantum coherence in light-harvesting systems
and have implications for quantum biology and quantum optics.

\begin{acknowledgement}
This work is supported by NSF (CHE 1800301 and CHE 1836913).
Pei-Yun Yang was partially supported by Ministry of Science and
Technology overseas project of Taiwan under Grant ID number 107-2917-I-564-011.
The key results of this paper were presented at the Banff meeting, August, 2019.
We thank Prof. Paul Brumer for the helpful discussions and for the exchange of recent manuscripts.
\end{acknowledgement}

\begin{suppinfo}
The Supporting Information is available: S1. Master Equation and Steady-state Solution; S2. Decomposition of Steady-state Coherence; S3. Energy Flux;
 S4. Average Trapping Time; S5. Delocalized Trapping; S6. Quantum Networks.
\end{suppinfo}

\newpage

\bibliography{bibfile}

\begin{mcitethebibliography}{49}
\bibitem[bea()]{beatings}
The molecular origin of 'quantum beats' in 2D spectroscopy has been debated in
  literature but is beyond the scope of this paper. The readers can find
  discussions in recent reviews. In this paper, we use excitonic coherence and
  Hamiltonian for the convenience of presentation, but the basic concept and
  analysis can be applied to any quantum states.\relax
\mciteBstWouldAddEndPunctfalse
\mciteSetBstMidEndSepPunct{\mcitedefaultmidpunct}
{}{\mcitedefaultseppunct}\relax
\EndOfBibitem
\end{mcitethebibliography}

\end{document}